\documentclass[copyright,creativecommons]{eptcs}
\providecommand{\event}{FORECAST 2016}

\usepackage[retainorgcmds]{IEEEtrantools}
\usepackage{amsmath} 
\usepackage{amsthm}
\usepackage{amssymb}
\usepackage{mathtools}
\usepackage{amsfonts}
\usepackage{todonotes}
\usepackage{titlesec}
\usepackage{tikz}
\usepackage{enumerate}
\usepackage{subcaption}
\usepackage{graphicx}
\usepackage{framed}
\usepackage{fancyhdr}
\usepackage{float}
\usepackage{subfloat}
\usepackage{listings}
\usepackage{float}
\usepackage{esvect}
\usepackage{bussproofs}
\usepackage{bbm}

\usetikzlibrary{matrix, automata, arrows, positioning, calc, decorations.pathreplacing}

\presetkeys{todonotes}{inline, size=\small}{}

\theoremstyle{plain}

\theoremstyle{definition}
\newtheorem{example}{Example}[section]
\newtheorem{definition}{Definition}[section]

\begingroup
    \makeatletter
    \@for\theoremstyle:=definition,remark,plain\do{%
        \expandafter\g@addto@macro\csname th@\theoremstyle\endcsname{%
            \addtolength\thm@preskip\parskip
            }%
        }
\endgroup

\makeatletter
\newbox\xrat@below
\newbox\xrat@above
\newcommand{\xratail}[2][]{%
    \setbox\xrat@below=\hbox{\ensuremath{\scriptscriptstyle #1}}%
    \setbox\xrat@above=\hbox{\ensuremath{\scriptscriptstyle #2}}%
    \pgfmathsetlengthmacro{\xrat@len}{max(\wd\xrat@below,\wd\xrat@above)+.8em}%
    \mathrel{\tikz[>->,baseline=-\the\dimexpr\fontdimen22\textfont2\relax]
                 \draw (0,0) -- node[below=-3pt] {\box\xrat@below}
                                node[above=-2pt] {\box\xrat@above}
                       (\xrat@len,0) ;}}
\makeatother

\makeatletter
\newbox\xrat@below
\newbox\xrat@above
\newcommand{\xratailb}[2][]{%
    \setbox\xrat@below=\hbox{\ensuremath{\scriptscriptstyle #1}}%
    \setbox\xrat@above=\hbox{\ensuremath{\scriptscriptstyle #2}}%
    \pgfmathsetlengthmacro{\xrat@len}{max(\wd\xrat@below,\wd\xrat@above)+.8em}%
    \mathrel{\tikz[>->,baseline=-\the\dimexpr\fontdimen22\textfont2\relax]
                 \draw (0,0) -- node[below=-3pt] {\box\xrat@below}
                                node[above=-3pt] {\box\xrat@above}
                       (\xrat@len,0) ;}}
\makeatother

\DeclarePairedDelimiter\abs{\lvert}{\rvert}%
\DeclarePairedDelimiter\norm{\lVert}{\rVert}%

\DeclareMathOperator{\Ir}{\mathbf{IR}}
\DeclareMathOperator{\Wt}{\mathbf{Wt}}
\DeclareMathOperator{\Prob}{\mathbf{Pr}}

\makeatletter
\let\oldabs\abs
\def\abs{\@ifstar{\oldabs}{\oldabs*}}

\let\oldnorm\norm
\def\norm{\@ifstar{\oldnorm}{\oldnorm*}}
\makeatother

\newcommand*\rel{\mathbin{\mathcal{R}}}

\newcommand*\parc{\mathbin{\mathcal{\|}}}

\newcommand{\blue}[1]{\textcolor{black}{#1}}

\newcommand{\red}[1]{\textcolor{black}{#1}}

\title{Stochastic and Spatial Equivalences for PALOMA}
\author{Paul Piho
\institute{LFCS, School of Informatics,\\ University of Edinburgh}
\email{paul.piho@ed.ac.uk}
\and
Jane Hillston
\institute{LFCS, School of Informatics,\\ University of Edinburgh}
\email{jane.hillston@ed.ac.uk}
}
\def\titlerunning{Stochastic and Spatial Equivalences for PALOMA}
\def\authorrunning{P. Piho and J. Hillston}

\begin{document}
\raggedbottom
\maketitle

\begin{abstract}
We concentrate our study on a recent process algebra -- PALOMA -- intended to capture interactions between spatially distributed agents, for example in collective adaptive systems.
\red{New} agent-based semantic rules for deriving the underlying continuous time Markov chain are given in terms of State to Function Labelled Transition Systems.
\red{Furthermore} we define a bisimulation with respect to an isometric transformation of space allowing us to compare PALOMA models with respect to their relative rather
than absolute locations.
\end{abstract}

\section{Introduction}

PALOMA (Process Algebra for Located Markovian Agents)~\cite{FengH14, FengHG16} is a novel stochastic process algebra which captures the behaviour of agents 
who are distributed in space and whose interactions are affected by their relative positions. 
This model can be thought to capture many modern systems where, for example, the range of communication may be limited for devices using wireless communication
technologies or some areas may be known ``dead zones" from which no communication is possible.
In this paper we consider what it means for two agents to be equivalent, taking into consideration both their behaviour and their location\red{, and develop the 
formal underpinnings to allow such equivalence to be rigorously studied.}

The notion of Markovian bisimulation has become standard for stochastic process algebras, but as we will discuss, applied naively this
approach to equivalence checking is too strong, leaving little opportunity for a notion of equivalence that is not isomorphism.  Instead here
we consider equivalence of a component within the context of a given system.  This supports the idea of being able to substitute one component,
perhaps with a more efficient implementation, for another within a given system even though they may not exhibit exactly the same behaviour
in arbitrary contexts.  Similarly, when we come to consider the spatial aspects of behaviour our notion of equivalence aims to capture
the relative positions of components, rather than their absolute locations.

In this brief paper we aim to give the intuition and ideas behind our bisimulation, without giving all the definitions.  The rest of the paper
is structured as follows.  In Section~\ref{sec:paloma} we give a brief introduction to the PALOMA modelling language, while the semantics
of the language is outlined in Section~\ref{sec:semantics}.  In Section~\ref{sec:equivalence} we discuss a notion of equivalence based on equivalent
relative positions and behaviours.  We present our conclusions and discuss further work in Section~\ref{sec:conc}.

\section{PALOMA language}
\label{sec:paloma}

In this section we give a brief introduction to PALOMA; the interested reader is referred to \cite{FengH14, FengHG16} for more details.   
The spatial distribution of agents is a key feature of PALOMA models and we assume that there exists a finite set of locations, $Loc$  and 
all agent expressions in PALOMA are parameterised by a location $\ell \in Loc$, indicating the current location of the agent.  

The grammar of the language is as follows:
\begin{small}
\begin{IEEEeqnarray*}{rCl}
    \pi &::=& !!(\alpha,r)@\Ir\{\vv{\ell}\} \enspace\mid\enspace ?? (\alpha, p)@\mathbf{Wt}\{w\} \enspace\mid\enspace !(\alpha, r)@\Ir\{\vv{\ell}\} \enspace\mid\enspace ?(\alpha, p)@\Prob\{q\}
    \enspace\mid\enspace (\alpha, r) \\
    S(\ell) &::=& \pi.S'(\ell') \enspace\mid\enspace S_1(\ell) + S_2(\ell) \enspace\mid\enspace C  \\
    P &::=& S(\ell) \enspace\mid\enspace P \mathbin{\|} P
\end{IEEEeqnarray*}
\end{small}%

\noindent 
The two-level grammar, defines individual agents $S(\ell)$, whose behaviours are specified by the
actions they can undertake, with possible alternatives, and model components $P$, which are
comprised of parallel compositions of agents. The behaviour of individual agents are given by actions of five distinct types:
\begin{description}
\item[Unicast output $!!(\alpha,r)@\Ir\{\vv{\ell}\}$:]
Unicast is for point-to-point communication between a pair of agents \red{and is included in the language to model contention for resources in systems}.
\red{Each unicast output message has a label, $\alpha$, and a rate $r$, that determines the rate at which the output is performed.
The message is sent to locations specified by the set $\vv{\ell} \in 2^{Loc}$ interpreted as the \emph{influence range}.}
Any agent located within that range, which enables the corresponding $\alpha$-labelled unicast input action, is eligible to receive the action --- \red{that is, the label $\alpha$ is 
used to identify agents that can communicate with each other.}
Unicast actions are \emph{blocking} meaning that the sending agent can only proceed when there is a eligible receiver.
\item[Unicast input $??(\alpha, p)@\mathbf{Wt}\{w\} $:]  Each eligible receiver of a \red{unicast message $\alpha$} must be located within the
specified influence range, and each will have an associated \emph{weight} $w$. 
The weights are used to define a probability distribution
over the eligible receivers, i.e.\ if there are $i$ potential receivers, each with weight $w_i$ and $W = \sum_i w_i$ then the probability 
that the $j$th agent receives the message is $w_j/W$.  Once the message is received the receiving agent may or may not act on the
message (reflecting message failure, corruption etc.) with the specified probability $p$ i.e.\ with probability $1-p$ the agent will not
act on the message received.  If this occurs the message is lost --- it is not the case that it is subsequently assigned to one of the other
eligible receivers.
\item[Broadcast output $!(\alpha, r)@\Ir\{\vv{\ell}\}$:] As its name suggests, a broadcast action allows its sender to influence multiple
other agents.  As with the unicast output action, a broadcast output \red{message labelled $\alpha$} is
sent with a specified influence range \red{$\vv{\ell}$} and at a specified rate $r$.  \emph{All}
agents \red{with broadcast input prefix on label $\alpha$} located within that range may receive the message.  Moreover the output proceeds regardless of whether there are any eligible receivers
so broadcast output is non-blocking for the sender.
\item[Broadcast input $?(\alpha, p)@\Prob\{q\}$:] Each eligible receiver of a broadcast \red{message $\alpha$} must be located within the specified
input range.  Each such agent has a likelihood of receiving the message, recorded in the probability $q$.  For example, agents
closer to the sender may be more likely to receive the message.  Each agent independently decides whether the broadcast
is received or not (Bernoulli trials).  As with unicast input, the receiving agent may or may not act on the
message with the specified probability $p$ i.e.\ with probability $1-p$ the agent will not
act on the message received. 
\item[Spontaneous action $(\alpha, r)$:] These actions do not represent a communication but rather an
individual action by the agent which may change the state of the agent, for example, its location.
These can also be thought of as broadcast output actions whose influence range is the empty set.
\end{description}
All rates are assumed to be parameters of exponential distributions, meaning that the underlying stochastic model of a PALOMA
model is a continuous time Markov chain (CTMC). 
\medskip

\begin{example}
    Consider agents $Transmitter$ and $Receiver$ such that 
        \begin{IEEEeqnarray*}{lCl}
            Receiver(\ell) &:=& ??(message, p)@\Wt\{v\}.Receiver(\ell) \\
            Transmitter(\ell) &:=& !!(message, r)@\Ir\{\vv{\ell}\}.Transmitter(\ell)
        \end{IEEEeqnarray*}
    \noindent where $\ell$ denotes the current location of the agent and $\vv{\ell}$ denotes a
    set of locations in the range of the unicast message emitted by action
    $message$. 
    In a system where no agent sends a $message$ agent $Receiver$ does not perform any action.
    On the other hand if there is a component, say $Tranmitter$, that
    outputs a $message$ and the location of $Receiver$ is in the
    influence range of the message then $Receiver$ performs $message$ with a rate
    dependent on the rate at which $Transmitter$ unicasts $message$ and the probability that $Receiver$ receives it.
    Similarly, if the component $Transmitter$ does not have a recipient for the $message$, it remains blocked and never
    performs an action.
    \end{example}

\subsection{Conditional exit rates and probabilities}

Notions of equivalence in process algebras, such as bisimulation \cite{Milner89}, are typically based on the idea
of a pair of agents each being able to match the behaviour of the other. In the case of stochastic process algebras
such as PEPA, not only the type of action but also the rates at which they occur must in some sense be matched \cite{Hillston96}.
In order to make similar definitions for PALOMA we need to define some auxiliary functions which, given a syntactic
expression, extract information about the rates and probabilities which may be exhibited by the term. Space limitations
do not allow us to present all of them here, but we present those for unicast, which is the most involved case, to give the
reader an impression of how we proceed.

Denote the set of all sequential components of PALOMA parametrised by their location by $\mathcal{C}_{S}$ and the set of
model components by $\mathcal{C}$.
Let the set of action labels be defined as $Lab$ and the set of action types as $Type = \{ !!, ??, !, ?, \cdot \}$, 
where the interpretation of the symbols is clear, corresponding to the action types discussed above.
Let $Act$ denote the set of all actions.
The actions in the set $Act = Type \times Lab$ are defined by their label and their type.
Let $\mathcal{A}$ be the set of all syntactically defined actions.
Define the function $\Pi_{Act}: \mathcal{A} \to Act$ as a projection returning the label of the action \red{with its type}, 
e.g.\  $\Pi_{Act}(??(\alpha, p)@\Wt\{v\}) = \mathbin{??}\alpha$.
Similarly define the projection $\Pi_{Lab}: Act \to Lab$ returning just the label of the action and the function 
$\Pi_{Type}:Act \to Type$ returning the type of an action.

Denote by $\Pi_{Loc}$ the function returning the set of locations spanned by a
model component.
\begin{IEEEeqnarray*}{c}
    \Pi_{Loc}(S_1(\ell_1) \parc \cdots \parc S_n(\ell_n)) = \bigcup_{i=1}^{n} \{\ell_i\}
\end{IEEEeqnarray*}
Note that in the case of sequential components $\Pi_{Loc}$ will result in a singleton set
--- the location of the sequential component.

Suppose $Sys = S_1(\ell_1) \parc \cdots \parc S_n(\ell_n) \in \mathcal{C}$ for $n \in \mathbb{N}^{+}$.
Let the function $\mathrm{seq}$ return the set of all sequential components of $Sys$ in a set of locations $L$.
\begin{IEEEeqnarray*}{c}
    \mathrm{seq}(Sys, L) = \{ S_i(\ell_i) \mid \Pi_{Loc}(S_i(\ell_i)) \in L\}
\end{IEEEeqnarray*}

\subsubsection{Context unaware definitions}

When we consider a PALOMA component in isolation we can use the syntax
to find the potential rate, weight or probability associated with  this component and a given action.
Similar functions are defined for each form of prefix.  
From the point of view of the originator of a unicast action, the important measure is the rate at which the 
action is preformed.

\begin{definition}
    For all $\alpha \in Lab$, $a \in \mathcal{A}$, $\vv{\ell} \in 2^{Loc}$, and $S \in \mathcal{C}_{S}$
    define the function $s_{\alpha}^{!!}$ returning the rate of a unicast output action labelled $\alpha$ as follows.
    \begin{IEEEeqnarray*}{lCl}
        s_{\alpha}^{!!}\left(!!(\beta,r)@\Ir\{\vv{\ell}\}.S(\ell)\right) & = & 
        \begin{cases}
            r & \mbox{for $\alpha = \beta$}\\
            0 & \mbox{otherwise}
        \end{cases} \\*
        \IEEEstrut
        s_{\alpha}^{!!}\left(a.S(\ell)\right) & = & \hspace{0.8em} 0 \quad \mbox{if $\Pi_{Type}(a) \neq \> !!$} \\
        s_{\alpha}^{!!}\left(S_1(\ell) + S_2(\ell)\right) & = & \hspace{0.8em} s_{\alpha}^{!!}(S_1(\ell)) + s_{\alpha}^{!!}(S_2(\ell))
    \end{IEEEeqnarray*}
\end{definition}

\begin{example}
Consider the following components
    \begin{IEEEeqnarray*}{lCl}
        Tester(\ell_0) &:=& (message, r).Tester(\ell_0) \\
        Transmitter(\ell_0) &:=& !!(message, r).Transmitter(\ell_0) \\
        Receiver(\ell_1) &:=& ??(message, p)@\Wt\{v\}.Receiver(\ell_1)
    \end{IEEEeqnarray*}
Based on these definitions we can find:
\begin{align*}
    s_{message}^{!!}(Tester(\ell_0) + Transmitter(\ell_0)) = 0 + r = r && s_{message}^{!!}(Receiver(\ell_1)) = 0
\end{align*}
\end{example}

The rest of the context unaware definitions are given in a similar vein and just extract necessary syntactic information from the
component definitions.
Specifically we define the following functions:
\begin{description}
    \item[Unicast influence range $\Pi_{UniIR}(S, \alpha)$:] Given that $S$ has a unicast output prefix with label $\alpha$, the function returns the influence range 
    of unicast message $\alpha$ defined in the prefix.
    Otherwise, the function returns the empty set $\emptyset$.

    \item[Weight function $w_{\alpha}(S)$:] For a sequential component $S$ the function $w_{\alpha}(S)$ is defined similarly to $s_{\alpha}^{!!}$ with 
    base case $w_{\alpha}\left(??(\alpha, p)@\Wt\{w\}.S\right) = w$.
    In addition we define the weight function over parallel compositions and sets of sequential components by summing over the weights for each sequential component in
    the parallel composition or set. 

    \item[Probability function  $p_{\alpha}^{??}(S)$:] This is again similar to $s_{\alpha}^{!!}$ with base case 
    $p_{\alpha}^{??}\left(??(\alpha, p)@\Wt\{w\}.S\right) = p$.
\end{description}

\begin{example}
    Consider the following sequential components.
    \begin{IEEEeqnarray*}{lCl}
        Transmitter(\ell_0) &:=& \,!!(message, r)@\Ir\{\vv{\ell}\}.Transmitter(\ell_0) \\
        Receiver1(\ell_1)   &:=& \,??(message, p)@\Wt\{w_{r1}\}.Receiver1(\ell_1) \\
        Receiver2(\ell_2)   &:=& \,??(message, q)@\Wt\{w_{r2}\}.Receiver2(\ell_2)
    \end{IEEEeqnarray*}
    For the system given by $Sys = Transmitter(\ell_0) \parc Receiver1(\ell_1) \parc Receiver2(\ell_2)$
    the weight for receiving a unicast message $message$ is calculated as
    \begin{IEEEeqnarray*}{l}
        w_{message}(Sys) = w_{message}(Transmitter(\ell_0) \parc Receiver1(\ell_1) \parc Receiver2(\ell_2))
        = w_{r1} + w_{r2}
    \end{IEEEeqnarray*}
\end{example}

\subsection{Context-aware conditional exit rates}
Unfortunately the syntactic information alone is not sufficient to determine the rate at which an action will be witnessed in a PALOMA system.  
The spatial aspect, as captured by the influence range, plays an important role in determining both which actions are possible and potentially their rates and probabilities.  
Thus we also define some context-dependent functions.

\begin{definition}
    Let $\alpha$ be an action label in $Lab$.
    Define the rate at which the component $S(\ell) \in \mathcal{C}_{S}$ is capable of unicasting a message labelled $\alpha$ to a location $\ell'$ as
    follows:
    \begin{IEEEeqnarray*}{lCl}
        u_{\alpha}(\ell', !!(\beta, r)@\Ir\{\vv{\ell}\}.S(\ell)) &=&
        \begin{cases} 
             s_{\alpha}^{!!}(!!(\beta, r)@\Ir\{\vv{\ell}\}.S(\ell)) & \mbox{if $\ell' \in \Pi_{UniIR}(S(\ell), \alpha)$ and $\alpha = \beta$} \\
             0 & \mbox{otherwise}
        \end{cases}\\
        u_{\alpha}(\ell', S_1(\ell) + S_2(\ell)) &=& \hspace{0.8em} u_{\alpha}(\ell', S_1(\ell)) + u_{\alpha}(\ell', S_2(\ell))
    \end{IEEEeqnarray*}
\end{definition}

\begin{definition}
    Suppose $P = S_1(\ell_1) \parc \dots \parc S_n(\ell_n) \in \mathcal{C}$ for $n \in \mathbb{N}^{+}$ is a model component with $S_i(\ell_i) \in
    \mathcal{C}_{S}$ for all $1 \leq i \leq n$.
    Let $Sys$ be any other system serving as context. 
    Let $u_\alpha(\ell, Sys, P)$ be the rate at which a model component $P$ unicasts a message labelled $\alpha$ to location $\ell$ in the context of $Sys$, defined as
    \begin{IEEEeqnarray*}{c}
        u_{\alpha}(\ell, Sys, P) = \sum_{S \in \mathrm{seq}(P)} u_{\alpha}(\ell, S)\times\mathbbm{1}_{>0}\{w_{\alpha}\left(\mathrm{seq}(Sys \parc P, \> \Pi_{UniIR}(S, \alpha))\right)\} \\
        \mbox{where } \mathbbm{1}_{>0}(x) = 
        \begin{cases} 
            1  & \mbox{$x > 0$} \\
            0  & \mbox{otherwise}
        \end{cases}
    \end{IEEEeqnarray*}
    \red{For each sequential component $S$ of $P$ we calculate the total weight over the components in the influence range of $S$. The indicator 
    function $\mathbbm{1}_{>0}$ is set to $1$ if this weight is greater than $0$ --- meaning there are eligible receivers in the influence range. 
    The rate at which $P$ unicasts a message $\alpha$ to location $\ell$ is then defined as the sum of rates at which each sequential component $S$ of $P$ is \emph{capable} of
    said unicast multiplied by the indicator function ensuring that the blocking nature of unicast is taken into account.}
\end{definition}

The next definition deals with determining the probability of a sequential
component receiving the unicast message.
\begin{definition}
    Let $S_1(\ell)$ and $S_2(\ell')$ be sequential components and $Sys \in \mathcal{C}$ any model component.
    Suppose $!!(\alpha, r)@\Ir\{\vv{\ell}\}.S_2'(\ell'')$ is a prefix guarded term in the expression of $S_2(\ell')$.
    Then we define the probability of $S_1(\ell)$ receiving a unicast message with label
    $\alpha$ from $S_2(\ell')$, when composed in parallel with $Sys$ and $S_2(\ell')$, to be: 
    \begin{IEEEeqnarray*}{l}
        p_{\alpha}(S_1(\ell), Sys, S_2(\ell')) =
        \begin{cases}
            \frac{w_{\alpha}(S_1(\ell))}{w_{\alpha}(\mathrm{seq}(Sys \parc S_1(\ell), \vv{\ell}))} & \mbox{if $\ell \in \vv{\ell}$} \\
            0 & \mbox{otherwise}
        \end{cases}
    \end{IEEEeqnarray*}
\end{definition}

Once similar definitions have been defined for broadcast and spontaneous actions we are in a position to define the context-aware exit rate.
\begin{definition}
    Let $Sys \in \mathcal{C}$ be a system in which the component $S \in
   \mathcal{C}_{S}$ appears.
    Let $a \in \mathcal{A}$ be any action with label $\alpha$.
    Define the \emph{context-aware exit rate} $R$ for agents by the following:
    \begin{IEEEeqnarray*}{rCl}
        R_{a}(Sys, S(\ell_0)) = 
            \begin{cases}
                s_{\alpha}(S(\ell_0)) & \text{if $\Pi_{Type}(a) = \mathbin{\cdot}$} \\
                b_{\alpha}(S(\ell_0)) & \mbox{if $\Pi_{Type}(a) = \> !$} \\
                b_{\alpha}(\ell_0, Sys) p_{\alpha}^{?}(S(\ell_0))& \mbox{if $\Pi_{Type}(a) = \>?$} \\
                \max\limits_{\ell \in \Pi_{Loc}(Sys)}\{u_{\alpha}(\ell, Sys,S(\ell_0))\} & \text{if $\Pi_{Type}(a) = \> !!$} \\
                \sum\limits_{T \in Seq(Sys)}u_{\alpha}(\ell_0, Sys, T) p_{\alpha}(S(\ell_0), Sys, T) p_{\alpha}^{??}(S(\ell_0)) & \text{if $\Pi_{Type}(a) = \> ??$} 
            \end{cases}
    \end{IEEEeqnarray*}
    \noindent Now consider a model component $P = S_1(\ell_1) \parc \dots \parc S_n(\ell_n)$ with $S_i(\ell_i) \in \mathcal{C}_{S}$ for all $1 \leq i \leq n$ ($n \in \mathbb{N}$) 
    and suppose it is a part of the system $Sys$.
    Then define
    \begin{IEEEeqnarray*}{rCl}
        R_{a}(Sys, P) = \sum_{i=1}^{n} R_{a}(Sys \parc (P \setminus S_i(\ell_i)), S_i(\ell_i))
    \end{IEEEeqnarray*}
    \noindent where $P \setminus S_i(\ell_i)$ denotes the model component $P$ with $S_i(\ell_i)$ removed.
\end{definition}

Finally we define the rate at which action $a \in Act$ is  performed over a set of locations.
\begin{definition}
 Consider a model component $P = S_1(\ell_1) \parc \dots \parc
    S_n(\ell_n)$ with $S_i(\ell_i) \in \mathcal{C}_{S}$ for all $1 \leq i \leq n$ ($n \in \mathbb{N}$) and suppose it is a part of the system $Sys$.
    Let $L$ be a set of locations of interest.
    We define $R_{a}(L, Sys, P)$, the rate at which action $a$ is performed by $P$ in locations $L$, within the context of system $Sys$ to be:
    \begin{IEEEeqnarray*}{rCl}
        R_{a}(L, Sys, P) = \sum_{S \> \in \> \mathrm{seq}(P, L)} R_{a}(Sys \parc (P \setminus S), S)
    \end{IEEEeqnarray*}
\end{definition}

\section{Semantics}
\label{sec:semantics}

The definition of the semantics of PALOMA will proceed in the FuTS (State to Function Labelled Transition Systems) framework as presented in~\cite{NicolaLLM13}.
In general, the transition rules in FuTSs are given as triplets $s \xratailb{\lambda} f$
where $s$ denotes a source state, $\lambda$ the label of the transition and $f$ the continuation function associating a value of suitable type
to each state $s'$.
The shorthand $\left[s_1 \mapsto v_1, \cdots, s_n \mapsto v_n \right]$ is used to denote a function $f$ such that $f(s_i) = v_i$ for $i = 1, \cdots n$.
This kind of functional treatment of transition rules is going to allow us to give more concise definitions of semantic rules as many possible branches of
model evolutions can be captured within a single rule.

In the case of PALOMA semantics we are going to define the set of states as the set of all model components $\mathcal{C}$.
For convenience, the treatment of semantic rules is split into two steps where the following types of transition relations are considered separately:
\begin{description}
    \item[Capability relation] Denoted by $s \xratailb{\lambda}_{c} f$ where $f: \mathcal{C} \to [0,1]$.
    The aim is to describe actions that a defined model component is capable of and introduce probabilities for all possible states resulting from the said action firing.
    For example, a component including a prefix for unicast input will be capable of the unicast input action firing with some probability dependent on the context.
    The function $f$ will assign a probability for possible continuation states.

    \item[Stochastic relation] Denoted by $s \xratailb{\lambda}_{s} f$ where $f: \mathcal{C} \to \mathbb{R}^{\geq 0}$.
    These rules are used to generate the CTMC and thus need to assign rates to each available transition.
\end{description}
As mentioned in Section~\ref{sec:paloma}, the calculation of rates of actions for each component depend the system they appear in (a PALOMA model
component) and thus we use $Sys$ as a place-holder for any such PALOMA model component serving as context.
In the following, we use $P_1 \equiv P_2$ to denote that model components $P_1$ and $P_2$ are syntactically equivalent.

\subsection{Capability relations}
The only capability relations of interest here are ones for broadcast and unicast input actions as these are the only ones that can either succeed or fail depending
on the rest of the context system $Sys$.

The labels $\lambda_c$, of the FuTSs rules are given by the following grammar where $\alpha \in \mathsf{Lab}$ denotes the action labels:
\begin{IEEEeqnarray*}{rCllll}
    \lambda_c ::= \> 
    &     & (?\alpha,  &\vv{\ell}, \> &Sys) \quad &\mbox{Broadcast input}  \\
    &\mid & (??\alpha, &\vv{\ell}, \> &Sys) \quad &\mbox{Unicast input} 
\end{IEEEeqnarray*}

The semantic rules given in Figure~\ref{fig:caprules} use the definitions from Section~\ref{sec:paloma} to extract necessary
information from the syntactic definitions of components.

The rules \textsf{BrIn} and \textsf{UniIn} are the primitive rules describing the capability of sequential components to perform a broadcast or unicast input action, 
respectively, given the set of locations $\vv{\ell}$ denoting the influence range of the message and a context system $Sys$.
In both cases the function $f$, which is defined over all states, gives the probability of a transition to each state given the action has fired.
For \textsf{BrIn} the calculation only depends on the parameters $p$ and $q$ given explicitly in the syntactic definition of the component.
For \textsf{UniIn} the likelihood of the component receiving the message, $\frac{w}{w_{\alpha}(Sys)}$, is calculated on the basis that there may be many eligible receivers of the
given message in $Sys$.

The rule \textsf{BrSystem} is used to deal with parallel compositions of model components that can act as broadcast message receivers.
\blue{
Note that the outcomes of all the broadcast input actions in a system are independent of each other. 
Thus the probability of $P_1 \parc P_2$ transitioning to $P_1' \parc P_2'$ due to a broadcast input action is the product of the
probabilities of $P_1$ and $P_2$ respectively making the corresponding transitions.}

For unicast input actions the rule \textsf{ParllelUniIn} is just saying that no two components can perform the unicast input on the same label simultaneously.

\begin{figure}[t!]%
    \scriptsize

    \begin{minipage}[c]{0.35\linewidth}
        \makebox[3em][l]{\textsf{BrIn}} \quad
        $?(\alpha, p)@\Prob\{q\}.S \xratail{\left(?\alpha, \vv{\ell}, \> Sys\right)}_{c} f$
    \end{minipage}
    \begin{minipage}[c]{0.3\linewidth}
        if $\Pi_{Loc}\left(?(\alpha, p)@\Prob\{q\}.S\right) \in \vv{\ell}$ 
    \end{minipage}
    \begin{minipage}[c]{0.4\linewidth}
        $f(s) = \begin{cases}
            pq & \mbox{if $s \equiv S$} \\
            1 - pq & \mbox{if $s \equiv \> ?(\alpha, p)@\Prob\{q\}.S$} \\
            0 & \mbox{otherwise}
        \end{cases}$
    \end{minipage}

    \bigskip

    \begin{minipage}[c]{0.35\linewidth}
        \makebox[3em][l]{\textsf{UniIn}} \quad
        $??(\alpha, p)@\Wt\{w\}.S \xratail{(??\alpha, \vv{\ell}, \> Sys)}_{c} f$   
    \end{minipage}
    \begin{minipage}[c]{0.3\linewidth}
        if $\Pi_{Loc}\left(??(\alpha, p)@\Wt\{w\}.S\right) \in \vv{\ell}$
    \end{minipage}
    \begin{minipage}[c]{0.4\linewidth}
        $f(s) = 
        \begin{cases}
            \dfrac{wp}{w_{\alpha}(Seq)} & \mbox{if $s \equiv S$} \\[1em]
            \dfrac{w(1-p)}{w_{\alpha}(Seq)} & \mbox{if $s \equiv \> ??(\alpha, p)@\Wt\{w\}.S$} \\[1em]
            0 & \mbox{otherwise}
        \end{cases}$\\

        \medskip

        where $Seq = \mathrm{seq}(Sys, \vv{\ell})$
    \end{minipage}

    \bigskip

    \begin{minipage}[c]{0.45\linewidth}
        \AxiomC{$P_1 \xratail{\left( ?\alpha, \vv{\ell}, \> Sys\right)}_{c} f_1 \quad P_2 \xratail{\left( ?\alpha, \vv{\ell}, \> Sys\right)}_{c} f_2$}
        \LeftLabel{\makebox[4em][l]{\textsf{BrSystem}} \quad}
        \UnaryInfC{$P_1 \parc P_2 \xratail{\left(?\alpha, \vv{\ell}, \> Sys\right)}_c g$}
        \DisplayProof
    \end{minipage}
    \begin{minipage}[c]{0.5\linewidth}
        \makebox[3cm]{}
        $g(s) =
        \begin{cases}
            f_{1}(P_1') f_{2}(P_2')  & \mbox{if $s \equiv P_1' \parc P_2'$} \\
            0 & \mbox{otherwise}
        \end{cases}$
    \end{minipage}

    \bigskip

    \begin{minipage}[c]{0.45\linewidth}
        \AxiomC{$S_1 \xratail{\left(??\alpha, \vv{\ell}, \> Sys\right)}_c f_1 \quad S_2 \xratail{\left(??\alpha, \vv{\ell}, \> Sys\right)}_c f_2$}
        \LeftLabel{\makebox[4em][l]{\textsf{ParllelUniIn}} \quad}
        \UnaryInfC{$S_1 \parc S_2 \xratail{\left(??\alpha, \vv{\ell}, \> Sys\right)}_c g$}
        \DisplayProof
    \end{minipage}
    \begin{minipage}[c]{0.5\linewidth}
        \makebox[3cm]{}
        $g(s) =
        \begin{cases}
            f_{1}(S_1') & \mbox{if $s \equiv S_1' \parc S_2$} \\
            f_{2}(S_2') & \mbox{if $s \equiv S_1 \parc S_2'$} \\
            0 & \mbox{otherwise}
        \end{cases}$
    \end{minipage}

    \bigskip

    \begin{minipage}[c]{0.25\linewidth}
        \AxiomC{$P_1 \xratail{\lambda_c}_{c} f$}
        \LeftLabel{\makebox[4em][l]{\textsf{Choice}} \quad}
        \UnaryInfC{$P_1 + P_2 \xratail{\lambda_c}_{c} f $}
        \DisplayProof
    \end{minipage}
    \begin{minipage}[c]{0.25\linewidth}
        \AxiomC{$P_2 \xratail{\lambda_c}_{c} f$}
        \UnaryInfC{$P_1 + P_2 \xratail{\lambda_c}_{c} f$}
        \DisplayProof
    \end{minipage}
    \red{
    \begin{minipage}[c]{0.5\textwidth}
        \makebox[2cm]{}
        \AxiomC{$P \xratail{\lambda_c}_{c} f \qquad X := P$}
        \LeftLabel{\makebox[1.5cm][l]{\textsf{Constant}} \quad}
        \UnaryInfC{$X \xratail{\lambda_c}_{c} f$}
        \DisplayProof
    \end{minipage}%
    }
    \caption{Capability rules for communication}
    \label{fig:caprules}
\end{figure}

\subsection{Stochastic relations}

Firstly we need to define a set of labels for stochastic relations.
It will be necessary to carry around the set of locations $\vv{\ell}$ in the labels to distinguish between actions having the same label and type but affecting a different set of
components due to their influence range.
In addition, including the system $Sys$ in the labels ensures that the communication rules are only applied to components in the same system.
The set of labels for stochastic relations is thus defined as follows:
\begin{IEEEeqnarray*}{rCllll}
    \lambda_s ::= \> 
    &     & (\alpha,   &\emptyset, \> &Sys) \quad &\mbox{Spontaneous action}\\
    &\mid & (!\alpha,  &\vv{\ell}, \> &Sys) \quad &\mbox{Broadcast communication}\\
    &\mid & (!!\alpha, &\vv{\ell}, \> &Sys) \quad &\mbox{Unicast communication}
\end{IEEEeqnarray*}

\red{The stochastic rules are summarised in Figure~\ref{fig:stochrules}.}
Firstly we have rules \textsf{Br}, \textsf{Uni} and \textsf{SpAct} that just define the primitive rules for all spontaneous actions and give the 
rates at which the defined transitions can happen.
For the rule \textsf{Uni} the side-condition is needed to ensure that there are eligible receivers available in the system. 

The rules \textsf{BrCombo} and \textsf{UniPair} are to combine the capability rules with stochastic rules to give rates of system state transitions that are induced
by broadcast or unicast message passing.
\textsf{BrCombo} takes as premise the existence of components $S$ and $P$ such that $S$ can perform the broadcast communication action defined by stochastic relations and
$P$ is capable of broadcast input.
The rate at which the parallel composition $S \parc P$ reaches the next state $S' \parc P'$ is given by the function $f \otimes g$ which is defined as the product of $f$ applied to the 
$S'$ and $g$ applied to $P$.
The unicast case is treated similarly.
\begin{figure}[hb!]
    \scriptsize
    \begin{subfigure}{\textwidth}
        \begin{minipage}{0.58\textwidth}
            \makebox[1.5cm][l]{\textsf{Br}} \quad
            $!(\alpha, r)@\Ir\{\vv{\ell}\}.S \xratail{\left(!\alpha, \vv{\ell}, \> Sys\right)}_{s} \left[S \mapsto r\right]$
        \end{minipage}
        
        \bigskip

        \begin{minipage}{\textwidth}
            \makebox[1.5cm][l]{\textsf{Uni}} \quad
            $!!(\alpha, r)@\Ir\{\vv{\ell}\}.S \xratail{\left(!!\alpha, \vv{\ell}, \> Sys\right)}_{s} \left[S \mapsto r\right]$
            \qquad if there exists $S$ such that $S \xratail{\left(??\alpha, \vv{\ell}, \> Sys\right)}_c f$
        \end{minipage}

        \bigskip

        \begin{minipage}{0.58\textwidth}
            \makebox[1.5cm][l]{\textsf{SpAct}} \quad
            $(\alpha, r).S \xratail{(\alpha, \emptyset, \> Sys)}_{s} \left[S \mapsto r\right]$
        \end{minipage}
        \caption{Primitive rules}
    \end{subfigure}

    \bigskip

    \begin{subfigure}{\textwidth}
        \begin{minipage}{0.58\textwidth}
            \AxiomC{$S \xratail{\left(!\alpha, \vv{\ell}, \> Sys\right)}_s f  \quad P \xratail{(?\alpha, \vv{\ell}, \> Sys)}_{c} g$}
            \LeftLabel{\makebox[1.5cm][l]{\textsf{BrCombo}} \quad}
            \UnaryInfC{$S \parc P \xratail{\left(!\alpha?, \vv{\ell}, \> Sys \right)} f \otimes g$}
            \DisplayProof
        \end{minipage}%
        \begin{minipage}{0.5\textwidth}
            $(f \otimes g)(s)= 
            \begin{cases}
                f(S')g(P') & \mbox{if $s \equiv S' \parc P'$}\\
                0 & \mbox{otherwise}
            \end{cases}$
        \end{minipage}

        \bigskip

        \begin{minipage}{0.58\textwidth}
            \AxiomC{$S \xratail{\left(!!\alpha, \vv{\ell}, \> Sys\right)}_{s} f \quad P \xratail{(??\alpha, \vv{\ell}, \> Sys)}_{c} g$}
            \LeftLabel{\makebox[1.5cm][l]{\textsf{UniPair}} \quad}
            \UnaryInfC{$S \parc P \xratail{\left(!!\alpha, \vv{\ell}, \> Sys\right)}_s f \otimes g $}
            \DisplayProof
        \end{minipage}
        \begin{minipage}{0.5\textwidth}
            $(f \otimes g)(s)= 
            \begin{cases}
                f(S')g(P') & \mbox{if $s \equiv S' \parc P'$}\\
                0 & \mbox{otherwise}
            \end{cases}$
        \end{minipage}
        \caption{Combining with capabilities}
    \end{subfigure}

    \bigskip 

    \begin{subfigure}{\textwidth}
        \begin{minipage}{0.58\textwidth}
            \AxiomC{$P_1 \xratail{\lambda_s}_{s} f$}
            \LeftLabel{\makebox[1.5cm][l]{\textsf{Parallel}} \quad}
            \UnaryInfC{$P_1 \parc P_2 \xratail{\lambda_s}_{s} f \otimes Id$}
            \DisplayProof
        \end{minipage}%
        \begin{minipage}{0.5\textwidth}
            $(f \otimes Id)(s) = 
            \begin{cases}
                f(P_1') & \mbox{if $s \equiv P_1' \parc P_2$} \\
                0 & \mbox{otherwise}
            \end{cases}$
        \end{minipage}

        \bigskip
        
        \begin{minipage}{0.58\textwidth}
            \AxiomC{$P_2 \xratail{\lambda_s}_{s} f$}
            \LeftLabel{\makebox[1.5cm][l]{} \quad}
            \UnaryInfC{$P_1 \parc P_2 \xratail{\lambda_s}_{s} Id \otimes f$}
            \DisplayProof
        \end{minipage}%
        \begin{minipage}{0.5\textwidth}
            $(Id \otimes f)(s) = 
            \begin{cases}
                f(P_2') & \mbox{if $s \equiv P_1 \parc P_2'$} \\
                0 & \mbox{otherwise}
            \end{cases}$
        \end{minipage}

        \bigskip 

        \begin{minipage}{0.15\textwidth}
            \AxiomC{$P_1 \xratail{\lambda_s}_{s} f$}
            \LeftLabel{\makebox[1.5cm][l]{\textsf{Choice}} \quad}
            \UnaryInfC{$P_1 + P_2 \xratail{\lambda_s}_{s} f$}
            \DisplayProof
        \end{minipage}%
        \begin{minipage}{0.43\textwidth}
            \AxiomC{$P_2 \xratail{\lambda_s}_{s} f$}
            \LeftLabel{\makebox[1.5cm][l]{} \quad}
            \UnaryInfC{$P_1 + P_2 \xratail{\lambda_s}_{s} f$}
            \DisplayProof
        \end{minipage}%
        \red{
        \begin{minipage}{0.15\textwidth}
            \AxiomC{$P \xratail{\lambda_s}_{s} f \qquad X := P$}
            \LeftLabel{\makebox[1.5cm][l]{\textsf{Constant}} \quad}
            \UnaryInfC{$X \xratail{\lambda_s}_{s} f$}
            \DisplayProof
        \end{minipage}%
        }
        \caption{Rules for composition}
    \end{subfigure}

    \caption{Stochastic rules for rates}
    \label{fig:stochrules}
\end{figure}

\blue{Suppose we want to derive a CTMC for the evolution of the model component $Sys$.
For that we need to consider all enabled stochastic transition rules from $Sys$.
The CTMC has a transition from the state $Sys$ to $Sys'$ if there is a transition $Sys \xratailb{\lambda_s}_s f$ such that
$f(Sys) \neq 0$. The next step is to consider all transitions from $Sys'$ and so on recursively until no new states are discovered and the full CTMC is generated.}

\section{Equivalence relations}
\label{sec:equivalence}

Firstly we will briefly cover a naive attempt to define a bisimulation on sequential components of PALOMA to demonstrate why it is not entirely trivial 
to deal with spatial properties of PALOMA models.
The approach that allows us to relax the conditions on spatial properties of defined models will be described in more detail.

In terms of semantic rules introduced in Section~\ref{sec:semantics} we are going to say that $S \xrightarrow{a} S'$ holds if there is a stochastic transition 
$Sys \xratailb{\lambda_s}_s f$ and a system $Sys'$ such that $S' \in \mathrm{seq}(Sys')$ and $f(Sys') \neq 0$.
In addition the label $\lambda_s$ is required to be such that
\begin{IEEEeqnarray*}{c}
    \lambda_s =
    \begin{cases}
        (\alpha, \emptyset, \> Sys) & \mbox{if $a = \alpha$} \\
        (!\alpha, \vv{\ell}, \> Sys) & \mbox{if $a = ?\alpha \lor !\alpha$} \\
        (!!\alpha, \vv{\ell}, \> Sys) & \mbox{if $a = ??\alpha \lor !!\alpha$} 
    \end{cases}
\end{IEEEeqnarray*}

As the behaviour of the PALOMA sequential component is parametrised by its location the natural interpretation would be to consider
locations as an inherent part of a component's state.  
This would lead to the following definition, making use of the syntax-derived rate function defined in Section~\ref{sec:paloma}.
\begin{definition}
\label{def:naive bisimulation}
    Let $Sys \in \mathcal{C}$ be any model component serving as a context.
    A binary relation $\mathcal{R}_{Sys}$ is a bisimulation over sequential components if, and only if, $(S(\ell_1), T(\ell_2)) \in \mathcal{R}_{Sys}$ implies, for all $a \in Act$
    \begin{enumerate}
        \item $R_{a}(Sys, S(\ell_1)) = R_{a}(Sys, T(\ell_2))$.
        
        \item $\ell_1 = \ell_2$.

        \item $S(\ell_1) \xrightarrow{a} S'(\ell_1')$ implies for some $T'(\ell_2')$, $T(\ell_2) \xrightarrow{a} T'(\ell_2')$ and $(S'(\ell_1'), T'(\ell_2')) \in \mathcal{R}_{Sys}$.

        \item $T(\ell_2) \xrightarrow{a} T'(\ell_2')$ implies for some $S'(\ell_1')$, $S(\ell_1) \xrightarrow{a} S'(\ell_1')$ and $(S'(\ell_1'), T'(\ell_2')) \in \mathcal{R}_{Sys}$.
    \end{enumerate}
 \end{definition}
    
This definition would give rise to an equivalence relation on PALOMA components with respect to the underlying context system.
However, Definition~\ref{def:naive bisimulation}  has some limitations due to the restrictive way in which location is treated, and we will not pursue it further.
Specifically, two sequential components which have identical behaviour in different locations will be considered non-equivalent in this setting.
This would lead to a very strict equivalence being defined on the model components of PALOMA\@.
A more interesting idea is to shift to considering relative locations between the sequential components. 
This will be explored in the following subsection.

\subsection{Relative locations}
\label{sec:isometry}

In order to consider relative locations between sequential components we need a notion of distance between the components.  Thus we consider the case where $Loc$ denotes a metric space.  Specifically we will consider the Euclidean plane
$\mathbb{R}^{2}$ \red{(extensions to different metric spaces are immediate)}.

The notion we make use of in the following discussion is that of isometries -- that is, maps between metric spaces that preserve the distances between points.
In particular we are interested in the set of Euclidean plane isometries of which we have four types: translations, rotations, reflections and glide reflections.
Denote the set of Euclidean plane isometries by $E(2)$.

The first definition we are going to give mimics the Definition~\ref{def:naive bisimulation} but allows the locations of the sequential components under consideration
to differ by an element in $E(2)$.

\begin{definition}
    Let $\phi \in E(2)$ and $Sys \in \mathcal{C}$ a system component serving as context.
    A binary relation $\mathcal{R}_{\phi, Sys}$ is a bisimulation with respect to $\phi$ over components if, and only if, $(S(\ell_1), T(\ell_2)) \in \mathcal{R}_{\phi, Sys}$ implies, for all $a \in Act$, that 
    \begin{enumerate}
        \item $R_{a}(Sys, S(\ell_1)) = R_{a}(Sys, T(\ell_2))$.

        \item $\phi(\ell_1) = \ell_2$.

        \item $S(\ell_1) \xrightarrow{a} S'(\ell_1')$ implies for some $T'(\ell_2')$, $T(\ell_2) \xrightarrow{a} T'(\ell_2')$ and $(S'(\ell_1'), T'(\ell_2')) \in \mathcal{R}_{\phi, Sys}$.

        \item $T(\ell_2) \xrightarrow{a} T'(\ell_2')$ implies for some $S'(\ell_1')$, $T(\ell_2) \xrightarrow{a} S'(\ell_1')$ and $(S'(\ell_1'), T'(\ell_2')) \in \mathcal{R}_{\phi, Sys}$.
    \end{enumerate}
\end{definition}

In this definition for sequential components the location plays little role.
The situation becomes more interesting when we attempt to extend the definition to model components $\mathcal{C}$ of PALOMA.

\begin{definition}\label{def:bisimulation wrt phi}
    Let $\phi \in E(2)$ and $Sys \in \mathcal{C}_{S}$ a model component serving as context.
    A binary relation $\mathcal{R}_{\phi, Sys}$ is a bisimulation with respect to $\phi$ over model components if, and only if, $(P, Q) \in \mathcal{R}_{\phi, Sys}$ implies, 
    for all $a \in Act$ and all sets of locations $L$
    \begin{enumerate}
        \item $R_{a}(L, Sys, P) = R_{a}(\phi(L), Sys, Q)$.

        \item $P \xrightarrow{a} P'$ implies for some $Q$, $Q \xrightarrow{a} Q'$ and $(P', Q') \in \mathcal{R}_{\phi, Sys}$.

        \item $Q \xrightarrow{a} Q'$ implies for some $P'$, $P \xrightarrow{a} P'$ and $(P', Q') \in \mathcal{R}_{\phi, Sys}$.
    \end{enumerate}
\end{definition}

From the definition we can easily see that any component is bisimilar to itself and that conditions are symmetric -- meaning we have 
$(P, Q) \in \mathcal{R}_{\phi, Sys} \implies (Q, P) \in \mathcal{R}_{\phi, Sys}$ -- and that transitivity holds.
To be able to define a bisimilarity as the largest bisimulation over the components would require us to verify that a union of bisimulations is again a bisimulation.
%

\begin{definition}\label{def:bisimilarity relation}
    Two model components $P_1, P_2$, defined over $\mathbb{R}^2$ are considered bisimilar with respect to context system $Sys$, denoted $P_1 \sim_{Sys} P_2$ 
    if there exists an isometry $\phi \in E(2)$ and a corresponding bisimulation $\rel_{\phi, Sys}$ such that $(P_1, P_2) \in \rel_{\phi, Sys}$.
\end{definition}
The simplest case we can consider is bisimilarity with respect to empty context system $Sys$ denoted by $\emptyset$.
We illustrate this in the following example.
\begin{example}
     \begin{IEEEeqnarray*}{llCl}
        &Transmitter(\ell_0)  &:=& !!(message\_move, r)@\Ir\{all\}.Transmitter(\ell_1) \\
        &Transmitter(\ell_1)  &:=& !!(message\_move, r)@\Ir\{all\}.Transmitter(\ell_0) \\
        &Receiver(\ell_1)     &:=& ??(message\_move, p)@\Wt\{v\}.Receiver(\ell_0) \\
        &Receiver(\ell_0)     &:=& ??(message\_move, q)@\Wt\{v\}.Receiver(\ell_1)
    \end{IEEEeqnarray*}
    For this example take $\ell_0 = (-1, 0)$ and $\ell_1 = (1, 0)$.
    The two systems we are going to analyse are
    \begin{IEEEeqnarray*}{l}
        Scenario_1 := Transmitter(\ell_0) \parc Receiver(\ell_1) \\
        Scenario_2 := Transmitter(\ell_1) \parc Receiver(\ell_0)
    \end{IEEEeqnarray*}
    It is clear that the systems are symmetric in the sense that if the locations in $Scenario_1$ are reflected along the $y$-axis we get $Scenario_2$.
    Denote the reflection along the $y$-axis as $\phi$.
    This give $\phi(\ell_0) = \ell_1$ and $\phi(\ell_1) = \ell_0$.
    
    It it intuitively clear that the two systems behave in the same way up to the starting location of the $Transmitter$ and $Receiver$ in both systems.
    Thus it makes sense to abstract away the absolute locations and consider the given systems observationally equivalent up to spatial transformation $\phi$.
    In the following we verify that applying Definition~\ref{def:bisimulation wrt phi} to these examples indeed agrees with the intuition.
    The two systems are considered on their own with no additional context -- that is the $Sys$ in Definiton~\ref{def:bisimulation wrt phi} becomes~$\emptyset$.
    \begin{IEEEeqnarray*}{l}
        R_{!!message\_move}(\ell_0 , \emptyset, Scenario_1) = r  \qquad \quad
        R_{??message\_move}(\ell_1 , \emptyset, Scenario_2) = rp
    \end{IEEEeqnarray*}
    and
    \begin{IEEEeqnarray*}{l}
        R_{!!message\_move}(\phi(\ell_0), \emptyset, Scenario_1) = R_{!!message\_move}(\ell_1, \emptyset, Scenario_1) = r \\
        R_{??message\_move}(\phi(\ell_1), \emptyset, Scenario_2 )= R_{!!message\_move}(\ell_0, \emptyset, Scenario_2) = rp 
    \end{IEEEeqnarray*}
    As the rest of the rates are $0$ then the first condition in Definition~\ref{def:bisimulation wrt phi} holds.
    To get the second and third conditions requires verifying that the rates also match for derivatives of the systems $Scenario_1$ and $Scenario_2$.
    This is not going to be done here but one can easily see that the same symmetries are going to hold throughout the evolution of the systems
    and thus the Definition~\ref{def:bisimilarity relation} would give that
    \begin{IEEEeqnarray*}{rCl}
        Scenario_1 \sim_{\emptyset} Scenario_2 
    \end{IEEEeqnarray*}
\end{example}

In the example we gave no additional context to the systems under study but the Definition~\ref{def:bisimulation wrt phi} allows for reasoning about the equivalence 
of the two components in the context of any given system.
The following example demonstrates that components being equivalent with respect to one context system does not imply equivalence with respect to other contexts.
\begin{example}
    \begin{IEEEeqnarray*}{llCl}
        &Transmitter(\ell_0)  &:=& !!(message, r)@\Ir\{all\}.Transmitter(\ell_0) \\
        &Receiver(\ell_0)     &:=& ??(message, r)@\Ir\{all\}.Receiver(\ell_0) \\
        &Sys                  &:=& Transmitter(\ell_0)
    \end{IEEEeqnarray*}
    It can be verified that according to Definition~\ref{def:bisimulation wrt phi} we have 
    \[
    Transmitter \sim_{\emptyset} Receiver
    \]
    as neither component can perform an action due to the blocking nature of unicast communication. 
    On the other hand we have
    \[
    Transmitter \not\sim_{Sys} Receiver
    \]
    as the system $Transmitter(\ell_0) \parc Transmitter(\ell_0)$ would not perform an action while in $Transmitter(\ell_0) \parc Receiver(\ell_0)$
    we have unicast communication happening.
\end{example}

\section{Conclusions and Future Work}
\label{sec:conc}
\blue{
The paper introducing the PALOMA language in its current form~\cite{FengHG16}
concentrated on the fluid analysis of CTMCs defined on population counts and gave semantic rules for generating a model in the Multi-message Multi-class Markovian 
Agents Model framework~\cite{CerottiGBCM10}.
}In order to have a rigorous foundation for bisimulation definitions we have introduced the new agent level semantics in the FuTSs framework~\cite{NicolaLLM13}.
\blue{Several other process algebras that capture the relative locations of interacting entities have been developed. 
In relation to systems biology there is, for example, SpacePi~\cite{johnEU08} where locations are defined in real coordinate spaces and for
wireless networks there is, for example, CWS~\cite{MezzettiS06} which makes no restrictions on the notion of location that can be used.  
However, there is very little work exploring notions of equivalence for spatially distributed systems.}

We presented an idea for a bisimulation of PALOMA models which allows us to abstract away explicitly defined locations of PALOMA components
and use relative locations of sequential components as the basis of the model comparison.
This idea relies on working over the Euclidean plane and being able to apply isometries to the model components of PALOMA leaving the relative spatial structure of
the model components intact.
As the behaviour of PALOMA components is dependent on the context in which they appear thus definitions of equivalences are given in terms of the context system.

The bisimulation ideas presented are intended to serve as a grounding for further development of model comparison and analysis methods for systems with explicitly defined
spatial location.
From the modelling and simulation perspective the aim of equivalence relations is to provide formal ways of reducing the state space of the underlying CTMC by allowing
us to swap out components in the model for ones generating a smaller state space while leaving the behaviour of the model the same up to some equivalence relation.
In particular, it is useful to consider such equivalence relations that induce a lumpable partition at the CTMC level.

\subsection*{Acknowledgements}
This work was supported by grant EP/L01503X/1 for the University of Edinburgh School of Informatics Centre for Doctoral Training in Pervasive Parallelism 
(http://pervasiveparallelism.inf.ed.ac.uk/) from the EPSRC, and by the EU-funded project, QUANTICOL 600708.

\bibliographystyle{eptcs}
\bibliography{bibliography}

\begin{thebibliography}{1}
\providecommand{\bibitemdeclare}[2]{}
\providecommand{\surnamestart}{}
\providecommand{\surnameend}{}
\providecommand{\urlprefix}{Available at }
\providecommand{\url}[1]{\texttt{#1}}
\providecommand{\href}[2]{\texttt{#2}}
\providecommand{\urlalt}[2]{\href{#1}{#2}}
\providecommand{\doi}[1]{doi:\urlalt{http://dx.doi.org/#1}{#1}}
\providecommand{\bibinfo}[2]{#2}

\bibitemdeclare{inproceedings}{CerottiGBCM10}
\bibitem{CerottiGBCM10}
\bibinfo{author}{Davide \surnamestart Cerotti\surnameend},
  \bibinfo{author}{Marco \surnamestart Gribaudo\surnameend},
  \bibinfo{author}{Andrea \surnamestart Bobbio\surnameend},
  \bibinfo{author}{Carlos Miguel~Tavares \surnamestart Calafate\surnameend} \&
  \bibinfo{author}{Pietro \surnamestart Manzoni\surnameend}
  (\bibinfo{year}{2010}): \emph{\bibinfo{title}{A Markovian Agent Model for
  Fire Propagation in Outdoor Environments}}.
\newblock In: {\sl \bibinfo{booktitle}{Computer Performance Engineering - 7th
  European Performance Engineering Workshop, {EPEW} 2010, Bertinoro, Italy,
  September 23-24, 2010. Proceedings}}, pp. \bibinfo{pages}{131--146},
  \doi{10.1007/978-3-642-15784-4\_9}.

\bibitemdeclare{article}{NicolaLLM13}
\bibitem{NicolaLLM13}
\bibinfo{author}{Rocco \surnamestart {De Nicola}\surnameend},
  \bibinfo{author}{Diego \surnamestart Latella\surnameend},
  \bibinfo{author}{Michele \surnamestart Loreti\surnameend} \&
  \bibinfo{author}{Mieke \surnamestart Massink\surnameend}
  (\bibinfo{year}{2013}): \emph{\bibinfo{title}{A uniform definition of
  stochastic process calculi}}.
\newblock {\sl \bibinfo{journal}{{ACM} Comput. Surv.}}
  \bibinfo{volume}{46}(\bibinfo{number}{1}), p.~\bibinfo{pages}{5},
  \doi{10.1145/2522968.2522973}.

\bibitemdeclare{inproceedings}{FengH14}
\bibitem{FengH14}
\bibinfo{author}{Cheng \surnamestart Feng\surnameend} \& \bibinfo{author}{Jane
  \surnamestart Hillston\surnameend} (\bibinfo{year}{2014}):
  \emph{\bibinfo{title}{{PALOMA:} {A} Process Algebra for Located Markovian
  Agents}}.
\newblock In: {\sl \bibinfo{booktitle}{Quantitative Evaluation of Systems -
  11th International Conference, {QEST} 2014, Florence, Italy, September 8-10,
  2014. Proceedings}}, pp. \bibinfo{pages}{265--280},
  \doi{10.1007/978-3-319-10696-0\_22}.

\bibitemdeclare{article}{FengHG16}
\bibitem{FengHG16}
\bibinfo{author}{Cheng \surnamestart Feng\surnameend}, \bibinfo{author}{Jane
  \surnamestart Hillston\surnameend} \& \bibinfo{author}{Vashti \surnamestart
  Galpin\surnameend} (\bibinfo{year}{2016}): \emph{\bibinfo{title}{Automatic
  Moment-Closure Approximation of Spatially Distributed Collective Adaptive
  Systems}}.
\newblock {\sl \bibinfo{journal}{{ACM} Trans. Model. Comput. Simul.}}
  \bibinfo{volume}{26}(\bibinfo{number}{4}), p.~\bibinfo{pages}{26},
  \doi{10.1145/2883608}.

\bibitemdeclare{book}{Hillston96}
\bibitem{Hillston96}
\bibinfo{author}{Jane \surnamestart Hillston\surnameend}
  (\bibinfo{year}{1996}): \emph{\bibinfo{title}{A Compositional Approach to
  Performance Modelling}}.
\newblock \bibinfo{publisher}{Cambridge University Press},
  \bibinfo{address}{New York, NY, USA},
  \doi{10.1017/CBO9780511569951}.

\bibitemdeclare{article}{johnEU08}
\bibitem{johnEU08}
\bibinfo{author}{Mathias \surnamestart John\surnameend},
  \bibinfo{author}{Roland \surnamestart Ewald\surnameend} \&
  \bibinfo{author}{Adelinde~M. \surnamestart Uhrmacher\surnameend}
  (\bibinfo{year}{2008}): \emph{\bibinfo{title}{A Spatial Extension to the pi
  Calculus}}.
\newblock {\sl \bibinfo{journal}{Electr. Notes Theor. Comput. Sci.}}
  \bibinfo{volume}{194}(\bibinfo{number}{3}), pp. \bibinfo{pages}{133--148},
  \doi{10.1016/j.entcs.2007.12.010}.

\bibitemdeclare{article}{MezzettiS06}
\bibitem{MezzettiS06}
\bibinfo{author}{Nicola \surnamestart Mezzetti\surnameend} \&
  \bibinfo{author}{Davide \surnamestart Sangiorgi\surnameend}
  (\bibinfo{year}{2006}): \emph{\bibinfo{title}{Towards a Calculus For Wireless
  Systems}}.
\newblock {\sl \bibinfo{journal}{Electr. Notes Theor. Comput. Sci.}}
  \bibinfo{volume}{158}, pp. \bibinfo{pages}{331--353},
  \doi{10.1016/j.entcs.2006.04.017}.

\bibitemdeclare{book}{Milner89}
\bibitem{Milner89}
\bibinfo{author}{Robin \surnamestart Milner\surnameend} (\bibinfo{year}{1989}):
  \emph{\bibinfo{title}{Communication and Concurrency}}.
\newblock \bibinfo{publisher}{Prentice-Hall, Inc.}, \bibinfo{address}{Upper
  Saddle River, NJ, USA}.

\end{thebibliography}

\end{document}